\documentclass{eage2021}
\usepackage{amsmath}
\usepackage{amsthm}
\usepackage{caption}
\usepackage{graphicx} 
\graphicspath{{./Figs/}} 
\usepackage{longtable} 
\DeclareCaptionFont{9pt}{\fontsize{9pt}{10pt}\selectfont}
\usepackage{fancyhdr} 
\usepackage{amsmath}
\usepackage{amsthm}
\usepackage{caption}
\usepackage{graphicx} 
\graphicspath{{./Figs/}} 
\usepackage{longtable} 
\DeclareCaptionFont{9pt}{\fontsize{9pt}{10pt}\selectfont}
\usepackage{fancyhdr} 
\usepackage{hyperref}
\usepackage{url}
\usepackage{amssymb}
\usepackage{xcolor}
\hypersetup{colorlinks,linkcolor={blue},citecolor={blue},urlcolor={red}}
\usepackage{caption}
\usepackage{subcaption}

\begin{document}
{\centerline{Large-scale finite-difference and finite-element frequency-domain seismic wave modeling}}
{\centerline{with multi-level domain-decomposition preconditioner}}
{\centerline{V. Dolean, P. Jolivet, P.-H. Tournier, L. Combe, S. Operto, S. Riffo}}


\section{Summary}
The emergence of long-offset sparse stationary-recording surveys carried out with ocean bottom nodes (OBN) makes frequency-domain full waveform inversion (FWI) attractive to manage compact volume of data and perform attenuation imaging. One challenge of frequency-domain FWI is the forward problem, which requires the solution of large and sparse linear systems with multiple right-hand sides. While direct methods are suitable for dense acquisitions and problems involving less than 100 million unknowns, iterative solver are more suitable for large computational domains covered by sparse OBN surveys. Here, we solve these linear systems with a Krylov subspace method preconditioned with the two-level Optimized Restricted Additive Schwarz (ORAS) domain decomposition preconditioner, the prefix optimized referring to the use of absorbing conditions at the subdomain interfaces. We implement this method with finite differences on uniform grid and finite elements on unstructured tetrahedral meshes. A simulation in a model where the velocity linearly increases with depth allows us to validate the accuracy of the two schemes against an analytical solution while highlighting how their relative cost varies with the band of propagated wavelengths. A simulation in the overthrust model  involving up to 2~billions of parameters allows us to tune the method and highlights its scalability.

\newpage
\clearpage

\section{Introduction}
\vspace{-0.5cm}
Most of the 3D Full Waveform Inversion (FWI) codes are developed today in the time domain because the time-marching forward modeling engines used to perform reverse time migration of towed-streamer data can be readily used for FWI. The emergence of ultra-long offset sparse stationary-recording acquisitions carried out with ocean bottom nodes (OBN) may put the frequency-domain formulation of FWI back in the spotlight because the inversion can be limited to a few frequencies for such acquisitions and attenuation can be straightforwardly taken into account in the forward and inverse problems. The main challenge of  frequency-domain FWI is related to the forward problem (the solution of the time-harmonic wave equation), which requires the solution of large, sparse and ill-conditioned linear systems with multiple right-hand sides for each frequency. Today, direct solver for sparse matrices likely provide the most efficient tools to solve these systems for dense seabed acquisitions and computational domains involving less than 100 millions unknowns \citep{Amestoy_2016_FFF}. For larger domains covered by sparse node arrays or for high frequency imaging, iterative solvers and domain decomposition preconditioner provide the replacement solution. While an up to date assessment of direct methods for 3D FWI is discussed in a companion study, we focus here on the development of an hybrid direct/iterative solver where the direct solver is used to solve the local problems in the subdomains of the preconditioner with incomplete Cholesky factorization. 
We develop the solver for a finite-element (FE) discretization on unstructured tetrahedral mesh with Lagrange elements of order 3 and an optimized 27-point finite difference (FD) scheme with adaptive coefficients \citep{Operto_2007_FDFD,Turkel_2013_CSO}. The iterative method relies on the Krylov subspace GMRES solver \citep{Saad_2003_IMS} with the Optimized Restricted Additive Schwarz (ORAS) domain decomposition preconditioner \citep{Graham:2017:RRD,Bonazzoli:2019:ADD}. Compared to preconditioners based upon shifted Laplacian \citep{Erlangga:2008:AIM}, ORAS is less sensitive to the shift (added attenuation) and can be used without it. 
In the following, we discuss the pros and cons of the two discretization methods with a numerical example. Then, we review the basic principles of the domain decomposition preconditioner before assessing the strong and weak scalability of the hybrid solver with the 3D SEG/EAGE Overthrust model and the FE discretization on unstructured mesh.

\vspace{-0.1cm}
\section{Tetrahedral Finite elements versus 27-point finite-differences}
\begin{figure}[ht!]
\begin{center}
\includegraphics[width=12cm,clip=true]{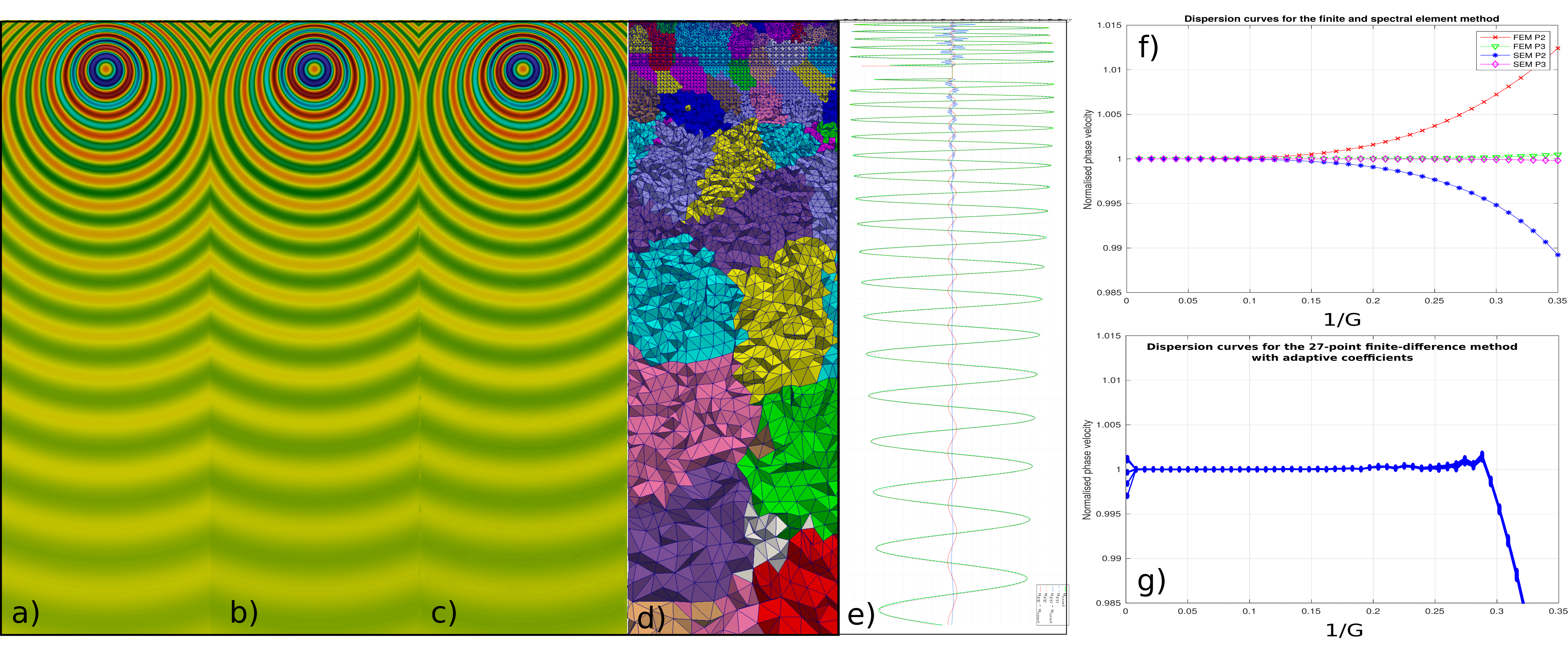}
\caption{FE and FD solutions against analytical solution. (a) FD solution. (b) Analytical (A) solution. (c) FE solution. (d) FE mesh. (e) Comparison between (a-c) along a vertical profile cross-cutting source position. Gain with offset is applied. FD: dash blue; FE: dash red; A: green. FD-A: blue. FE-A: red.  (f-g) Phase velocity dispersion curves for (f) FE and (g) FD. $G$: number of {\it{ppwl}}.}
\label{fig_df_ef}
\end{center}
\end{figure}

\vspace{-0.3cm}
The simplest mathematical model of acoustic wave propagation is the Helmholtz equation
\begin{equation}
\left(\Delta + k^2(\bold{x}) \right) u(\bold{x},\omega) = b(\bold{x},\omega), ~ \text{in a subsurface domain } \Omega,
\label{eqh}
\end{equation}
where $u$ is the monochromatic pressure wavefield, $b$ the source, $k(\bold{x},\omega)=\omega/c(\bold{x})$, with $\omega$ denoting frequency, $c(\bold{x})$ the wavespeed (which is complex valued in viscous media) and $\bold{x}=(x,y,z) \in \Omega$. \\
After discretization, eq. (\ref{eqh}) can be written in matrix form as
\begin{equation}
\bold{A} \bold{u} = \bold{b}.
\label{eq1}
\end{equation}
We implement the above equation with absorbing boundary conditions along the vertical and bottom faces of $\Omega$ and a homogeneous Dirichlet condition on the pressure along the top face. \\
We consider two discretizations of \eqref{eqh}. 
The first relies on the 27-point FD stencil in which compact $2^{nd}$-order accurate stencils minimize the numerical bandwidth and maximizes the sparsity of $\bold{A}$ while reaching a high-order accuracy by mixing consistent mass and stiffness matrices on different (rotated) coordinate systems \citep{Operto_2007_FDFD}. These sparsity and compactness properties are useful to minimize the matrix fill-in induced by a sparse direct solver, which is used to solve the local problems in each subdomains of the preconditioner (see next section).
The stiffness and  consistent mass matrices are weighted by coefficients that are computed by least-squares minimization of the numerical dispersion and anisotropy. Generally, the same coefficients are used in each row of the matrix. In this study, we implement adaptive coefficients in $\bold{A}$ that are matched to the local wavespeed to optimize the accuracy of the stencil in heterogeneous media \citep{Turkel_2013_CSO}.
The second relies on Lagrange finite elements on a tetrahedral mesh $\Gamma$ of the domain  $\Omega$. Compared to the 27-point FD method on uniform Cartesian grid, unstructured meshes in FE methods allow for the adaptation of the size of the elements to the local wavelength (the so-called $h$-adaptivity) and the conformal representation of complex known boundaries (topography, bathymetry). The drawback of the FE discretization is the higher number of degrees of freedom per element for a given accuracy.
A numerical and dispersion analysis of the FE method lead to the conclusion that Polynomials of degree $3$ (P3) are necessary to reach a sufficient accuracy for a discretization of four points per wavelength ({\it{ppwl}}) \citep{Dolean_2020_LSL}, which is further supported by the dispersion analysis of \citet{Ainsworth:2010:OBS} (Fig~\ref{fig_df_ef}). This discretization  is typically used for FWI application because it is the coarsest one allowing to sample an heterogeneity of size half a wavelength. The phase velocity dispersion curves for the 27-point FD stencil compare well with those obtained for the P3 finite elements (Fig.~\ref{fig_df_ef}).
To gain a first-hand understanding of the relative cost and accuracy of the two methods, we perform a simulation in a 3D medium of size 2~km $\times$ 4~km $\times$ 12~km  where the velocity linearly increases with depth ($c(x,y,z)=c_0 + \alpha \times z$) with $c_0$=1~km/s. We validate the numerical solutions against an analytical solution \citep{Kuvshinov_2006_EST} using $\alpha$=0.8 and $\alpha$=2 (Tab.~\ref{tab_df_ef}). Source is at 1~km depth and frequency is 8~Hz. The grid interval in the FD grid ($h$=31.25~m) corresponds to 4~{\it{ppwl}}, while the size of the tetrahedral elements is matched to the local wavelength for the FE simulation with a discretization rule of 3~{\it{ppwl}}. Fig.~\ref{fig_df_ef}(a-e) highlights the high accuracy achieved by the two discretizations. The number of degrees of freedom ($\#dof$) and the error involved in the two simulations highlight how the discretization method should be selected based on the specifications of the application in terms of size, structural complexity and dynamic of propagated wavelengths $\lambda$ (Tab.~\ref{tab_df_ef}).
\begin{table}[ht!]
\begin{center}
\begin{tabular}{|c|c|c|c|c|c|c|}
\hline
& &  & \multicolumn{2}{|c|}{$\#$dof (M)} & \multicolumn{2}{|c|}{$\ell{2}$ error}  \\ \hline
$\alpha(s^{-1})$    & $\lambda_{min}(m)$ & $\lambda_{max}(m)$ & FD  & FE & FD  & FE  \\ \hline
0.8 & 125 &  1200 & 13 &  28 &  0.0079 &  0.034 \\ \hline
2  & 125 &  3125 & 13 &  16 & 0.044 &  0.034 \\ \hline
\end{tabular}
\end{center}
\caption{FD {\it{.vs.}} FE. $\alpha$: gradient of the wavespeed. $\lambda_{min,max}(m)$: Min./max. wavelength. $\#$dof: number of degrees of freedom. $\ell{2}$ error: Least-squares error {\it{.wrt.}} analytical solution.}
\label{tab_df_ef}
\end{table}
\vspace{-0.5cm}
%
%
\vspace{-0.6cm}
\section{Domain decomposition preconditioner}
\vspace{-0.3cm}
We now review the preconditioner that we use to solve efficiently the linear system (\ref{eq1}).
A well-known iterative solver for this type of indefinite linear systems is the Krylov subspace Generalized Minimal RESidual Method (GMRES) \citep{Saad_2003_IMS}. However, the Helmholtz operator requires efficient preconditioning which can be done by domain decomposition \citep[][ section 2.2.1]{Dolean:2015:IDD}. \\
In this study, we solve system \eqref{eq1} with a two-level domain decomposition preconditioner $\bold{M}^{-1}$
\begin{equation}
\label{2lvl}
\bold{M}^{-1}  = \bold{M}^{-1}_1 (I - \bold{A} \bold{Q}) + \bold{Q}, \quad \text{with } \bold{Q} = \bold{Z} \bold{E}^{-1} \bold{Z}^T, \quad \bold{E} = \bold{Z}^T \bold{A} \bold{Z},  \\
\end{equation}
where $\bold{M}^{-1}_1$ is the one-level domain decomposition preconditioner called Optimized Restricted Additive Schwarz (ORAS) and $\bold{Z}^T$ is the interpolation matrix from the FE space defined on $\Gamma$ onto a FE space defined on a coarse mesh $\Gamma_H$. The construction of the domain decomposition preconditioner is described in detail in \citet{Bonazzoli:2019:ADD}. Let $\left\{\Gamma_i\right\}_{1 \le i \le N_d}$ be an overlapping decomposition of the mesh $\Gamma$ into $N_d$ subdomains. Let $\left\{\bold{A}_i\right\}_{1 \le i \le N_d}$ denote local Helmholtz operators with absorbing (or transmission) boundary conditions at the subdomain interfaces. The one-level ORAS preconditioner is
\begin{equation}
\bold{M}^{-1}_1 = \sum_{i=1}^{N_d} \bold{R}_i^T \bold{D}_i \bold{A}_i^{-1} \bold{R}_i,
\label{oras}
\end{equation}
where $\left\{\bold{R}_i\right\}_{1 \le i \le N_d}$ are the Boolean restriction matrices from the global to the local finite element spaces and $\left\{\bold{D}_i\right\}_{1 \le i \le N_d}$ are local diagonal matrices representing the partition of unity.\\
The key ingredient of the ORAS method is that the local matrices $\left\{\bold{A}_i\right\}_{1 \le i \le N_d}$ incorporate more efficient boundary conditions (i.e. absorbing boundary conditions) than in the standard RAS preconditioner based on local Dirichlet boundary value problems.
The coarse problem $\bold{E}$ in~\eqref{2lvl} is also solved iteratively by performing $10$ GMRES iterations with a one-level ORAS preconditioner. We use the same spatial subdomain partitioning for the coarse and fine meshes. Each computing core is assigned to one spatial subdomain and holds the corresponding coarse and fine local matrices. Each application of the global preconditioner $\bold{M}^{-1}$ relies on local concurrent subdomain solves on the coarse and fine levels, which are performed by a direct solver. This hybrid direct/iterative solver requires careful strong scalability analysis to achieve the best compromise between parallel efficiency and memory storage.\\
%
\vspace{-0.7cm}
\section{Numerical results}
\begin{figure}[ht!]
\begin{center}
\includegraphics[width=16cm]{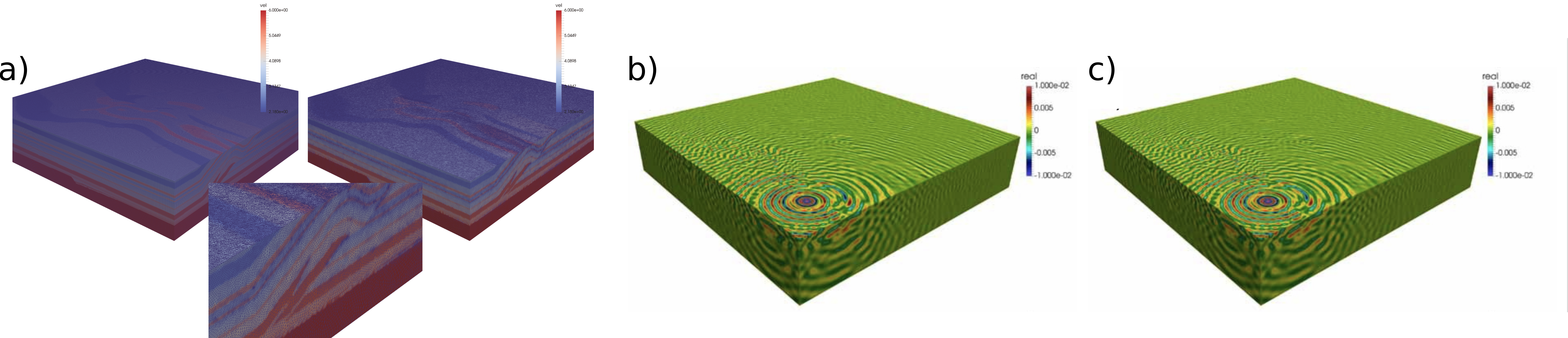}
\caption{(a) Tetrahedral meshing of Overthrust. Solutions on (b) regular and  (c) adaptive meshes.}
\label{over1}
\end{center}
\end{figure}

\vspace{-0.3cm}
The two-level solver is implemented using the high-performance domain decomposition library HPDMM (\url{http://github.com/hpddm/hpddm}) (High-Performance unified framework for Domain Decomposition Methods)  \citep{Jolivet:2013:SDD}.
We assess the solver on the Ir\`ene supercomputer of TGCC (\url{http://www-hpc.cea.fr}) with the 3D $20 \times 20 \times 4.65$ km SEG/EAGE Overthrust model \citep{Aminzadeh_1997_DSO}. We perform wave simulation with P3 finite elements on regular and adaptive tetrahedral meshes (Fig. \ref{over1}a) for the 5~Hz, 10~Hz and 20~Hz frequencies (Tab.~\ref{tab_over}) in double and single precision. The average length of the element edges is set to 5 nodes per minimum wavelength on the regular tetrahedral mesh, and 5 nodes per local wavelengths in the adaptive tetrahedral mesh (2.5 for the coarser mesh used in the two-level method).
We use a homogeneous Dirichlet boundary condition at the surface and first-order absorbing boundary conditions along the other five faces of the model. The source is located at (2.5,2.5,0.58) km.  For weak scalability analysis, we keep $\#$dofs per subdomain roughly constant from one frequency to the next (Tab.~\ref{tab_over}). The $h$-adaptivity in the unstructured tetrahedral mesh decreases $\#$dofs relative to the regular mesh by a factor of 2.07. The stopping tolerance $\epsilon$ for GMRES is set to $10^{-3}$. The consistency between the 10~Hz wavefields computed in the regular and adaptive tetrahedral meshes is shown in Fig.~\ref{over1}(b-c). 
First, we carry out a set of numerical simulations at 5~Hz on the regular mesh in order to illustrate the benefits of performing computations in single precision arithmetic (versus double precision), as well as using an approximate factorization for the fine local matrices to apply $\bold{A}_i^{-1}$ in~\eqref{oras}. More precisely, we compare incomplete Cholesky factorization (ICC) to complete Cholesky factorization performed by Intel MKL PARDISO. The experiments are performed on 1060 cores with P3 finite elements and 5 ppwl, resulting in 74 million dofs. Results are reported in Tab.~\ref{tab_prec}. First, we can see that performing the whole computation in single precision instead of double precision yields a speedup of about 1.4 for the solution phase. The number of GMRES iterations is the same, there is no loss of accuracy or additional numerical instability. Additionally, the setup phase is drastically reduced (speedup 1.8) when performing Cholesky factorization in single precision. Second, we can see that using an incomplete Cholesky factorization for the fine local matrices yields a speedup of about 1.6 with respect to complete factorization, once again with no effect on the number of GMRES iterations. Moreover, the memory savings are pretty significant: with complete Cholesky factorization we run out of memory with 768 cores, while the simulation runs on 265 cores using ICC.
In the rest of this paper, the experiments are performed in single precision and using incomplete Cholesky factorization for the fine local matrices. Timings for the adaptive tetrahedral mesh are around two times smaller than those obtained on the regular mesh (Tab.~\ref{tab_over}). The simulation at 20~Hz on the adaptive mesh involves 2,285 millions of dofs and requires 16,960 cores. The elapsed time achieved by the 2-level preconditioner is 15s and 37s for 10~Hz and 20~Hz respectively (Tab.~\ref{tab_over}).
%
%
\vspace{-0.2cm}
\section{Conclusions}
\vspace{-0.5cm}
We propose a highly-scalable hybrid direct/iterative solver based upon a domain decomposition preconditioner as a forward engine to perform large-scale 3D frequency domain FWI of sparse stationary-recording acquisitions. The method is implemented with a finite-difference and finite-element method to select the most suitable scheme for the case study at hand. This forward engine should be used when the size of the problem outreaches the capability of leading-edge sparse direct solvers. \\

%
\vspace{-0.4cm}
\textbf{Acknowledgments:}\\
This study was granted access to the HPC resources of SIGAMM (\url{http://crimson.oca.eu}) and CINES/IDRIS under the allocation 0596 made by GENCI. This study was partially funded by the WIND consortium (\url{https://www.geoazur.fr/WIND}) sponsored by Chevron, Shell and Total.
\begin{table}[ht!]
\begin{center}
\begin{tabular}{|c|c|c|c|c|}
\hline
\multicolumn{5}{|c|}{\bf{Cartesian grid, \; f = 5Hz}} \\ \hline
precision & fine local solver & $\#$it & setup(s) &  $T_{GM}(s)$ \\ \hline
double & Cholesky & 10 & 92.5 & 15.5 \\ \hline
double & ICC & 10 & 30.2 & 8.9 \\ \hline
single & Cholesky & 10 & 50.3 & 10.3 \\ \hline
single & ICC & 10 & 25.8 & 6.3 \\ \hline
\end{tabular}
\end{center}
\caption{Comparison between Cholesky and incomplete Cholesky factorizations (ICC) of local matrices at the fine level, and single {\it{.vs.}} double precision arithmetic for the whole computation, at 5Hz with P3 FEs and 5 ppwl (74M dofs) on 1060 cores. $\#$it: iteration count. setup: Elapsed time for the setup phase (assembly and factorization of local matrices).$T_{GM}$: Elapsed time in GMRES ($\epsilon=10^{-3}$).}
\label{tab_prec}
\end{table}
\vspace{-0.4cm}
\begin{table}[ht!]
\begin{center}
\begin{tabular}{|c|c|c|c|c|c|}
\hline
\multicolumn{6}{|c|}{\bf{Regular tetrahedral mesh}} \\ \hline
f(Hz) & $\#$core & $\#$elts  (M)& $\#$dofs (M)   & $\#$it & $T_{GM}(s)$ \\ \hline
{\bf{5}} & 265 & 16 & 74 & 7 & 16s \\ \hline
{\bf{10}} & 2,120 & 131 & 575 & 15 & 33s \\ \hline
\multicolumn{6}{|c|}{\bf{Adaptive tetrahedral mesh}} \\ \hline
f(Hz) & $\#$core & $\#$elts  (M)& $\#$dofs (M)   & $\#$it & $T_{GM}(s)$ \\ \hline
{\bf{10}} & 2,120 & 63 & 286 & 14 & 15s \\ \hline
{\bf{20}} & 16,960 & 506 & 2,285 & 30 & 37s \\ \hline
\end{tabular}
\end{center}
\caption{Simulation cost in regular/adaptive meshes. $f$: frequency; $\#$core: number of cores; $\#$elts: number of elements; $\#$dofs: number of dofs; $\#$it: iteration count. $T_{GM}$: Elapsed time in GMRES.}
\label{tab_over}
\end{table}
\vspace{-0.5cm}
\small{
\newcommand{\SortNoop}[1]{}

}


\end{document}